\documentclass[aps,twocolumn,preprintnumbers,amsmath,amssymb,
a4paper,showkeys,letterpaper,preprintnumbers]{revtex4}
\usepackage{graphicx}
\usepackage{dcolumn}
\usepackage{bm}

\begin{document}

\preprint{BI--TP 2007/04}

\title{Correlating anomalies of the microwave sky: The Good, the Evil
and the Axis} 

\author{Aleksandar Raki\'c and Dominik J. Schwarz}
\affiliation{Fakult\"at f\"ur Physik, Universit\"at Bielefeld,
Postfach 100131, D--33501 Bielefeld, Germany} 
\date{\today}

\begin{abstract}
At the largest angular scales the presence of a number of unexpected 
features has been confirmed by the latest measurements of the cosmic
microwave background (CMB). Among them are the anomalous
alignment of the quadrupole and octopole with each other as well as
the stubborn lack of angular correlation on scales
$>60^\circ$. We search for correlations between these two
phenomena and demonstrate their absence. A Monte Carlo 
likelihood analysis confirms previous studies and shows that the 
joint likelihood of both anomalies is incompatible with the 
best-fit $\Lambda$ Cold Dark Matter model at $>99.95\%$~C.L. 
Extending also to higher multipoles, a common special direction 
(the `Axis of Evil') has been identified. In the seek for an 
explanation of the anomalies, several studies invoke effects 
that exhibit an axial symmetry. We find that this interpretation of
the `Axis of Evil'
is inconsistent with three-year data from the Wilkinson Microwave 
Anisotropy Probe (WMAP).
The data require a preferred plane, whereupon the axis
is just the normal direction. Rotational symmetry 
within that plane is ruled out at high confidence.
\end{abstract}


\maketitle

\section{Standpoint}

With the emergence of more and more precise and detailed
cosmological observations the inflationary $\Lambda$ cold dark matter 
($\Lambda$CDM) model remains to provide a surprisingly good fit.
Thereby the most precise and distinguished lever arm is
provided by the cosmic microwave background (CMB). The standard
inflationary model predicts approximately scale-invariant,
statistically isotropic and gaussian temperature fluctuations on the
surface of last scattering and is fully consistent with the data.
But after the release of three years of mission data from the 
Wilkinson Microwave Anisotropy Probe (WMAP) \cite{wmap_all,lambda} 
there remain at least open questions and at most serious
challenges upon the inflationary $\Lambda$CDM model of cosmology. 

Based on the high precision measurements of WMAP, 
a couple of anomalies on the microwave sky have been identified.
These anomalies manifest themselves at the largest angular scales, 
mainly among the quadrupole and octopole (the dipole is overwhelmingly
dominated 
by our local motion with respect to the CMB), but also extending 
to somewhat higher multipoles. The corresponding anomalies may be 
divided into two types:

First, and already seen by the Cosmic
Background Explorer's Differential Microwave Radiometer (COBE-DMR) \cite{cobe}
and confirmed by the first-year analysis of the WMAP team \cite{spergel03}, 
there is a lack of angular two-point correlation on scales between
$60^\circ$ and $170^\circ$ in all wavebands. In \cite{copi06} the
angular two-point correlation function of the three-year WMAP
measurements has been computed. Going form COBE-DMR to WMAP(3yr) the
lack of correlation persists and moreover it has been outlined
\cite{copi06} that among the two-point angular correlation functions
none of the almost vanishing cut-sky wavebands matches the
reconstructed full sky and neither one of the latter
matches the prediction of the best-fit $\Lambda$CDM model. 
This disagreement has been shown to be even more
distinctive in the WMAP(3yr) data than in the WMAP(1yr) data and is
found to be unexpected at $99\%$~C.L.\ with respect to the three-year
Internal Linear Combination [ILC(3yr)] cut-sky. Recently, it has 
been shown \cite{hajian} that indeed quadrupole and octopole are  
responsible for the lack of correlation and that most of the 
large-scale angular power comes from two distinct regions within 
the galactic plane (only $9\%$ of the sky).

Second, there exist anomalies concerning the phase relationships of
the quadrupole and octopole. There are a number of remarkable
\textit{alignment anomalies} \cite{deOliv03,schwarz04}, e.g.~an
unexpected alignment of the quadrupole and octopole with the dipole
and with the equinox at $99.7\%$~C.L.\ and $99.8\%$~C.L., respectively
\cite{copi06}. In contrast to such \textit{extrinsic} alignments, that
is alignments of the low multipoles with some physical direction or plane, like
the dipole or the ecliptic, the \textit{intrinsic} alignment between
quadrupole and octopole does not know about external directions. In
this work, we address the intrinsic alignment of quadrupole and
octopole with each other, which from the ILC(3yr) map is found 
to be anomalous at $99.6\%$~C.L.\  with respect to the expectation for 
an statistically isotropic and gaussian sky \cite{copi06}.

Both types of CMB phenomena challenge the statement of statistical
isotropy of the CMB sky at largest angular scales. Here we want to
study the relation between the lack of angular correlation and
intrinsic alignment of quadrupole and octopole.

In \cite{aoe05} it has been shown that intrinsic alignments among 
multipole moments extend also to higher ones and it has been 
proposed that the strange alignments at large angular scales involve a 
preferred direction, called the `Axis of Evil'.  This axis points approximately
towards $(l,b)\simeq(-100^\circ,60^\circ)$ and is
identified as the direction where several low multipoles ($\ell=2-5$)
are dominated by one $m$-mode when the multipole frame is rotated
into the direction of the axis. Recently, \cite{aoe06} the analysis of
the `Axis of Evil' has been redone in the light of the WMAP(3yr) with
the use of Bayesian techniques \cite{ocam06}.
It was argued \cite{deOliv06} that the `Axis of Evil' is rather robust
against foreground contaminations and galactic cuts. A recent \cite{rassat06}
cross-correlation of CMB data and galaxy survey data shows
no evidence for an `Axis of Evil' in the observed large scale
structure. In contrast, recently an opposite claim has been put
forward \cite{longo}.

Motivated by these observed CMB anomalies, several mechanisms 
based on some axisymmetric effect have been proposed,
although the operational definition of the `Axis of Evil'
\cite{aoe05,aoe06} does  
not necessarily imply the existence of such a strong symmetry.
Among the various effects that have been suggested to possibly 
introduce a preferred axis into cosmology are: a spontaneous breaking of
statistical isotropy \cite{huhu06}, parity violation in general
relativity \cite{alex06}, anisotropic perturbations of dark energy 
\cite{mota,battye06}, residual large-scale anisotropies after inflation 
\cite{campa06,conta06}, or a primordial preferred direction \cite{ackerman07}.
At the same time it has been studied
\cite{rse1,rse2, silk} how the local Rees-Sciama 
effect \cite{rees68} of an extended foreground, non-linear in density contrast, 
affects the low multipole moments of the CMB via its time-varying 
gravitational potential. In a scenario with a
single  
overdensity the coefficients of the spherical harmonic decomposition,
the $a_{\ell m}$, become modified by only \textit{zonal harmonics}, 
i.e. $m=0$ modes. This is equivalent to an axial effect along the 
line connecting our position with the centre of the source. 

In fact, the observed pattern in the CMB for quadrupole and
octopole  is a nearly pure $a_{\ell \ell}$ mode respectively; as seen
in a frame where the ${\boldsymbol z}$-axis equals the normal of the plane defined
by the two quadrupole multipole vectors \cite{copi05}.
In \cite{copi06} it has been argued already, that foreground mechanisms
originating from a relatively small patch of the sky  would
mainly excite zonal modes. Moreover all additive effects where
extra contributions are added on top of the primordial fluctuations
would have difficulties explaining the low multipole power at large scales without a
chance cancellation.

In this work we study how the inclusion of a preferred axis compares with
the intrinsic multipole anomalies at largest scales. Our analysis is
restricted to axisymmetric effects on top of the primordial
fluctuations from standard inflation, thus secondary or systematic
effects. We quantify how poorly an axisymmetric effect at low
multipoles of whatever origin matches the three year-data of WMAP. We
demonstrate that there is no correlation between the two types of
intrinsic low-$\ell$ anomalies: the two-point correlation  
deficit and intrinsic alignment; and that there remains none even when
a preferred axis is introduced to the problem.      

The paper is organized as follows. In the next section the necessary
technical framework is introduced, 
especially the multipole vector formalism. Thereafter in
Sec.~\ref{sec3} we recapitulate the best-fit standard inflationary model (`the Good')
and its quantitative predictions for the low-$\ell$ microwave
sky as well as the anomalous (`the Evil') measurements of
WMAP(3yr). Before we conclude in Sec.~\ref{sec5}, the
consequences of an axial symmetry (`the Axis') on the
low-$\ell$ CMB sky are discussed in Sec.~\ref{sec4}.

\section{Choice of Statistic}
\label{sec2}

\begin{figure*}[ht!]
\includegraphics[width=3.46in,angle=0]{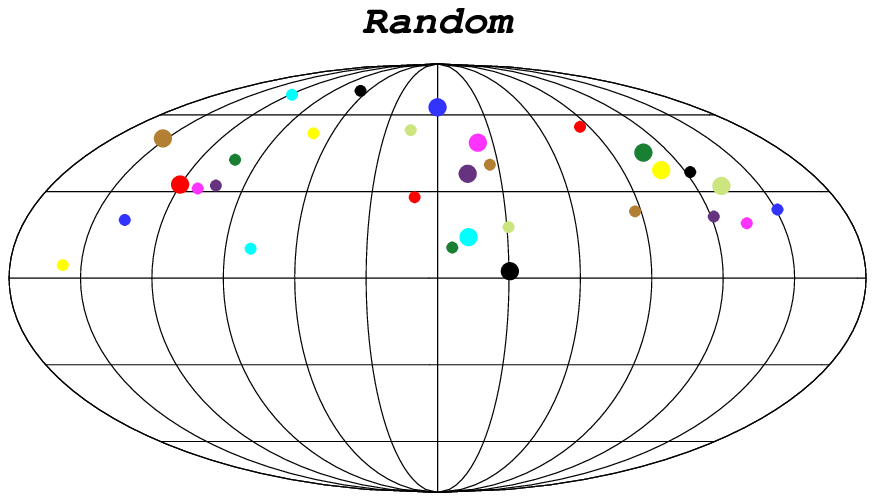}
\hspace*{-5mm}
\includegraphics[width=3.46in,angle=0]{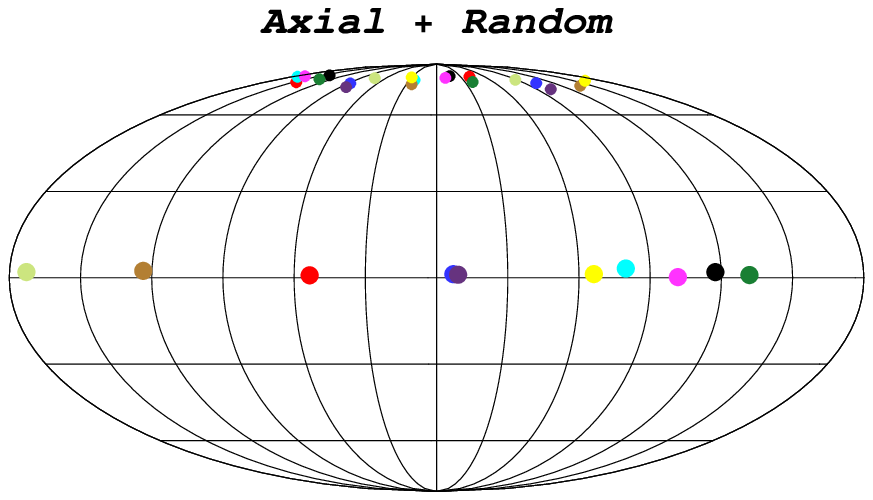}\\[4pt]
\includegraphics[width=3.1in,angle=0]{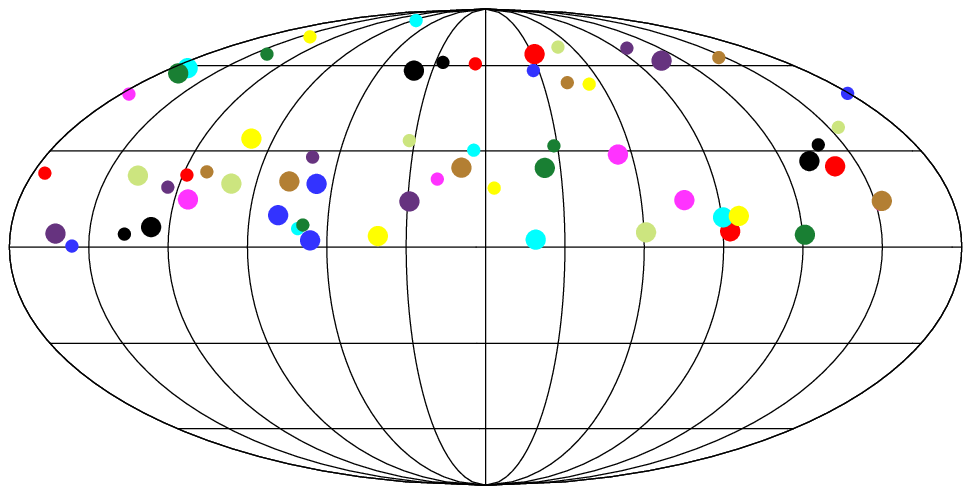}
\hspace*{4mm}
\includegraphics[width=3.1in,angle=0]{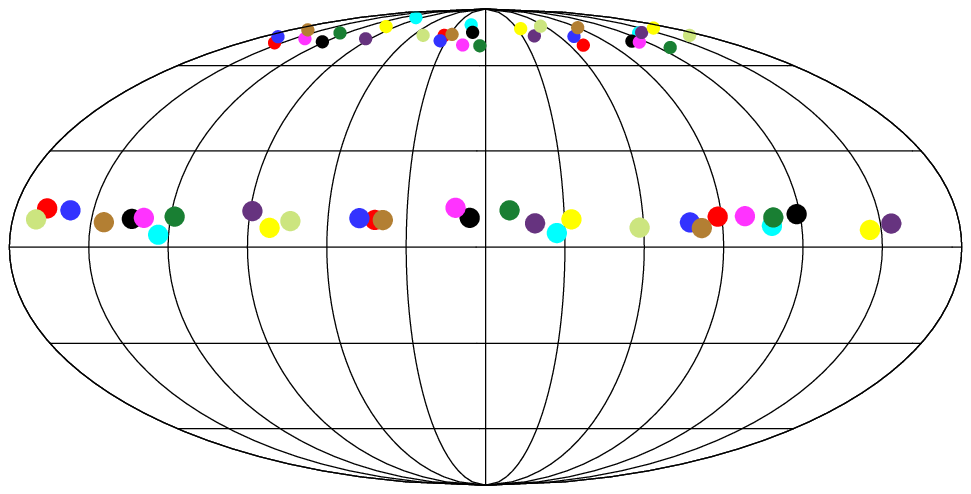}
  \caption{Mollweide projection of the sky with quadrupole (upper row)
and octopole (lower row) multipole vectors [Equation
(\ref{eq_multp})]. The mesh consists of steps in $30^\circ$. Displayed
are ten pairs of quadrupole vectors (small dots) and their ten area
vectors [Equation (\ref{eq_areav}) (big dots)] as well as ten triples
of octopole vectors (small dots) and their area vectors (big dots);
togetherness is indicated by color. The arbitrary sign of the vectors
has been used to gauge them all to the northern hemisphere. The
statistically isotropic and gaussian case (left column) is broken by
the imprint of a strong axial effect $a_{\ell 0}=1000\mu$K (right
column) whereupon multipole vectors move to the pole and area vectors
move to the equatorial plane. The onset of the shown separation of
multipole vectors and cross products can already be observed at
moderate axial contributions of $a_{\ell 0}\sim100\mu$K.}
\label{fig_mollw} 
\end{figure*}

A common observable is the multipole power. 
According to the standard perception of inflationary cosmology, the 
CMB fluctuations are believed to follow a gaussian statistic and to
be distributed in a statistically isotropic way.
The notion of statistical isotropy means that the expectation value of
pairs of coefficients $\langle a^*_{\ell^\prime m^\prime} a_{\ell m}
\rangle$ is proportional to $\delta_{\ell^\prime \ell} \,
\delta_{m^\prime m}$. Usually the proportionality constant measuring the
expectation value of the multipole on the full sky is estimated  by
\begin{equation}
 C_\ell \equiv \frac{1}{2\ell + 1} \sum_{m=-\ell}^\ell |a_{\ell m}|^2 = 
 \frac{1}{2\ell + 1} \int {\rm d}\Omega \;
T^2_\ell(\theta, \varphi) \; ,
\label{eq_cl}
\end{equation}
with $T_\ell$ being the $\ell$-th multipole of the CMB temperature
anisotropy. It can be expanded with the help of spherical harmonics
as:
$T_\ell = \sum_m a_{\ell m} Y_{\ell m}$. Note that, since we consider
multipole moments that are real, the $a_{\ell m}$ 
must fulfill an additional condition: $a^*_{\ell m} = (-1)^m a_{\ell -m}$.
Using the estimator (\ref{eq_cl}) the angular two-point correlation
function is given by
\begin{equation}
 C(\theta) = \frac{1}{4\pi} \sum_{\ell=0}^\infty (2\ell + 1) C_\ell 
P_\ell(\cos\theta) \; ,
\label{eq_2pt}
\end{equation}
where the $P_\ell$ are the Legendre Polynomials of $\ell$-th order. 

Besides of the multipole power itself, it is useful to introduce
an all-sky quantity that embraces all scales.  
As inspired by the $S_{1/2}$ statistic, presented in \cite{spergel03} for
measuring the lack of angular power at scales larger than
$60^\circ$, we use here an analogous all-sky statistic \cite{copi06}: 
\begin{equation}
 S_{\rm full} \equiv \int_{-1}^1 C^2(\theta) \, {\rm d (cos}\theta)
\;.
\label{eq_sful}
\end{equation}
This is a measure of the total power squared on the full-sky.
In contrast to the $S_{1/2}$ statistic \cite{spergel03} the $S_{\rm
full}$ statistic does not contain any a priori knowledge on the
variation of the two point angular correlation (\ref{eq_2pt})
for angles $> 60^\circ$.
Here we are considering especially the large angular scales but we are not
interested in the monopole and dipole and thus arrive at
\begin{equation}
 S_{\rm full}^{\rm trunc} = \frac{1}{8\pi^2} \left( 5C_2^2 + 7C_3^2 \right) \; .
\label{eq_sfull} 
\end{equation}
Of course all multipoles have to be considered for the full-sky
statistic (\ref{eq_sful}) but we use the truncated part
(\ref{eq_sfull}), because here the anomalies are most pronounced and we
want to check for the interplay of this part of the full-sky power
statistic with the other (phase) anomalies within quadrupole and octopole.
This part is then simply to be added to the rest of the sum of
(squared) multipole power in (\ref{eq_sful}), recovering the expression
for the full-sky. 

Next we turn to the statistics involving the phase relationships of
multipoles. We use the concept of Maxwell's multipole vectors
\cite{maxwell} in order to probe  
statistical isotropy, since this representation proved to be useful
for analyses of geometric alignments and special directions on the CMB sky. 
Normally the CMB data is decomposed into spherical harmonics
and the coefficients $a_{\ell m}$ containing the physics. Alternatively, 
with the use of the multipole vectors formalism \cite{copi04} we can
expand any real  
temperature multipole function on a sphere into  
\begin{eqnarray}
 T_\ell(\theta, \varphi) &=& \sum_{m=-\ell}^\ell a_{\ell m} Y_{\ell
m}(\theta, \varphi) \nonumber \\
&=& A^{(\ell)} 
 \left[ \, \prod_{i=1}^\ell \left( \hat{\boldsymbol v}^{(\ell,i)}
\cdot 
 \hat{\boldsymbol e}(\theta, \varphi) \right) -
\mathcal{L}_{\ell}(\theta, \varphi) \, 
 \right] \; . \nonumber \\
\label{eq_multp}
\end{eqnarray}
Therein $\hat{\boldsymbol v}^{(\ell,i)}$ is the $i$-th vector belonging
to the multipole $\ell$ and $\hat{\boldsymbol e}(\theta, \varphi) 
= (\sin\theta\cos\varphi , \sin\theta\sin\varphi , \cos\theta)$ stands
for a radial unit vector. With this decomposition all of the 
information in a  multipole moment is reorganized such that
we obtain a unique factorisation into  a scalar $A^{(\ell)}$ counting  
the total power and
$\ell$ unit vectors $\hat{\boldsymbol v}^{(\ell,i)}$ encoding all the
directional information. The product expansion term alone contains
also terms with `angular momentum' $\ell-2, \ell-4,...$ These residuals
are subtracted with the help of the term
$\mathcal{L}_{\ell}(\theta,\varphi)$. Because all the signs of the
multipole vectors may be absorbed in the scalar quantity $A^{(\ell)}$ the
multipole vectors are unique up to their sign which carries no
physical meaning. For definiteness, we define them to point to the
northern hemisphere.

In order to disclose correlations among the multipole vectors we first 
consider for each $\ell$ the $\ell(\ell-1)/2$ independent \textit{oriented areas}
built from the cross products \cite{schwarz04,copi05,copi06}: 
\begin{equation}
  {\boldsymbol w}^{(\ell;i,j)} \equiv \pm \; \hat{{\boldsymbol v}}^{(\ell,i)}
   \times \hat{{\boldsymbol v}}^{(\ell,j)} \; ,
\label{eq_areav}
\end{equation}
whereof we will also use the normalized vectors ${\boldsymbol n}^{(\ell;i,j)} 
\equiv {\boldsymbol w}^{(\ell;i,j)}/|{\boldsymbol w}^{(\ell;i,j)}|$. Now, in
\cite{schwarz04} and subsequent works the dot products of the area
vectors proved to be a handy expression in order to quantify
alignments of the multipole vectors among  
each other and also with external directions (which we do not consider here). 
The following measure as stated in \cite{weeks} and used in
\cite{schwarz04,copi05,copi06} serves as a natural choice
of a statistic in order to quantify the intrinsic alignment  
of quadrupole and octopole oriented areas:
\begin{equation}
 S_{{\boldsymbol w}{\boldsymbol w}} \; \equiv \; \frac{1}{3} \,  
 \sum_{i < j} \left| {\boldsymbol w}^{(2;1,2)} \cdot {\boldsymbol
w}^{(3;i,j)} \right| \, .  
\label{eq_sww}
\end{equation}
Note that we consider only the very largest scales, i.e. we use the statistic 
only for $\ell=2,3$. Analogously, a statistic involving the normalized area
vectors is given by:
\begin{equation}
 S_{{\boldsymbol n}{\boldsymbol n}} \; \equiv \; \frac{1}{3} \,  
 \sum_{i < j} \left| {\boldsymbol n}^{(2;1,2)} \cdot {\boldsymbol
n}^{(3;i,j)} \right| \, .
\label{eq_snn}  
\end{equation}

\begin{figure*}[ht!]
\includegraphics[height=3.4in,angle=270]{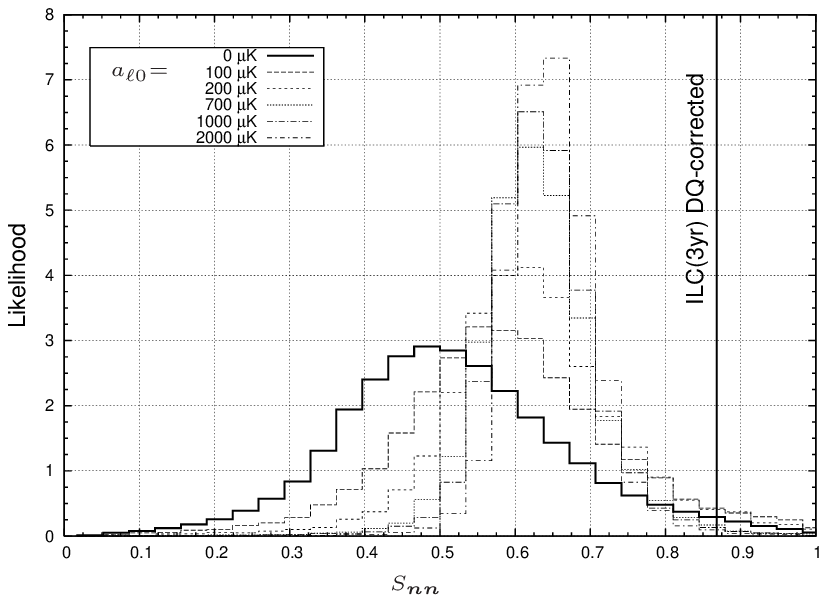}
\hspace*{8pt}
\includegraphics[height=3.4in,angle=270]{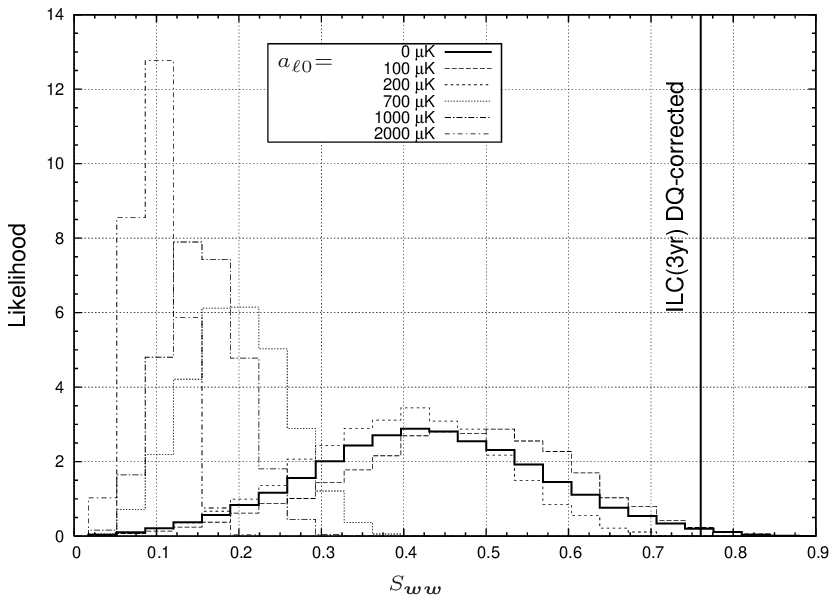}
\caption{Evolution of the Monte Carlo likelihood of the alignment
statistics $S_{{\boldsymbol n}{\boldsymbol n}}$ (\ref{eq_sww}) and
$S_{{\boldsymbol w}{\boldsymbol w}}$ (\ref{eq_snn}). The effect of an
axis in the CMB is modeled via increasing additional zonal harmonics
with coefficients $a_{\ell 0}$. At $a_{\ell 0}=1000\mu$K the
multipoles become purely zonal in good approximation. Regarding WMAP's
ILC(3yr) map $S_{{\boldsymbol n}{\boldsymbol n}}$ is unexpected at
$98.3\%$C.L.~and $S_{{\boldsymbol w}{\boldsymbol w}}$ is odd at
$99.5\%$C.L.~with respect to the statistically isotropic and gaussian
sky (bold histograms). The best improvement is reached for both
statistics at roughly $a_{\ell 0}=100\mu$K.}
\label{fig_Snn_Sww}
\end{figure*}

\section{Inflationary ${\bf \Lambda}$CDM predictions}
\label{sec3}

Standard inflationary $\Lambda$CDM cosmology requires the CMB
anisotropies to be gaussian and statistically isotropic.
For the subsequent analysis we have produced Monte Carlo realisations of the
harmonic coefficients $a_{\ell m}$ following the underlying
$\Lambda$CDM theory. From \cite{copi04} an algorithm
is available which we use to obtain Monte Carlo multipole vectors from
the coefficients. 
Mollweide maps of a sample
of random gaussian and statistically isotropic quadrupole and
octopole vectors as well as their normals are given in Figure
\ref{fig_mollw} (left column).

Concerning the question of correlations between the multipole power and
the alignment of multipole vectors, it appears natural to expect that
there is none. That is because we invoked gaussian random and
statistically isotropic skies leading to multipole vectors (\ref{eq_multp})
independent of the multipole power (\ref{eq_cl}). This assumption
needs to be tested and quantified. 

Nevertheless, a small correlation could be expected from the following
reason: Considering only multipoles up to some limiting power, the
resulting probability density distribution for the $a_{\ell m}$ 
must be non-gaussian. In fact, this restriction leads to a
negative kurtosis for the $a_{\ell m}$ distribution (the skewness
vanishes). Having that in mind, it appears suddenly unclear whether
the naive expectation of vanishing correlation of power with intrinsic
alignment will hold. Below we substantiate the absence of 
correlations by means of a Monte Carlo analysis.

Let us first look at the alignment anomalies. In Figure
\ref{fig_Snn_Sww} the likelihood of the quadrupole and octopole 
alignment statistics $S_{{\boldsymbol w}{\boldsymbol w}}$ and
$S_{{\boldsymbol n}{\boldsymbol n}}$ is shown. The predictions of the
standard inflationary $\Lambda$CDM model are shown as the bold
histograms respectively ($=$ vanishing axial contamination). According
to the three-year ILC map from WMAP
\cite{lambda} we get the following measured values for the alignment
statistics:
\begin{equation}
 S_{{\boldsymbol n}{\boldsymbol n}}^{\rm ILC(3yr)} = 0.8682
\;\;\;\;\;\;,\;\;\;\;\;\; 
 S_{{\boldsymbol w}{\boldsymbol w}}^{\rm ILC(3yr)} = 0.7604 \nonumber
\end{equation}
when \cite{copi06} corrected for the Doppler-quadrupole. The total
number of Monte Carlos we produced per sample is $N=10^5$. We
infer that the unmodified inflationary $\Lambda$CDM prediction is
unexpected at $98.3\%$ C.L.~with the $S_{{\boldsymbol n}{\boldsymbol
n}}$ statistic and unexpected at $99.5\%$ C.L.~\footnote{The value
quoted above was \cite{copi06} $99.6\%$ C.L. The small difference is 
due to the incorporation of the WMAP pixel noise in the Monte
Carlo analysis in \cite{copi06}.} with respect to the 
$S_{{\boldsymbol w}{\boldsymbol w}}$ statistic.

Next we consider the cross-correlation between the
intrinsic phase anomalies and the multipole power (\ref{eq_cl}) within
the low-$\ell$. For this we chose those $a_{\ell m}$ that allow for
say the lowest 
possible $5\%$ in the left tail of the distributions for $C_2$
\textit{and} $C_3$ that follow from statistical isotropy,
gaussianity and the $\Lambda$CDM best-fit to the WMAP data.
Then we compute the expression $S_{{\boldsymbol w}{\boldsymbol
w}}$ for the selected $a_{\ell m}$ and compare it to the according
ILC(3yr) value. As expected, no correlation is found, that is neither
the shape nor the expectation value of 
the alignment statistic is shifted. We find the same also for the
combination of the lowest allowed $5\%$ in $C_2$ and the highest $5\%$
from the right tail of the distribution of $C_3$ and the remaining two
possible combinations thereof.
As we do not find any correlations, we can conclude that the
$S_{{\boldsymbol w}{\boldsymbol w}}$ and $S_{{\boldsymbol n}{\boldsymbol n}}$
statistics are not sensitive to the non-gaussianity induced by the
restriction to low multipole power.

Moreover, we probe the opposite
direction by tagging those $a_{\ell m}$ that lie in the allowed right tail
of the $S_{{\boldsymbol w}{\boldsymbol w}}$ distribution with respect
to $S_{{\boldsymbol w}{\boldsymbol w}}^{\rm ILC(3yr)}$. The
\textit{distribution} of the multipole power for $C_2$ and $C_3$ made
of these $a_{\ell m}$ remains unchanged. 
The latter finding confirms that multipole power and the shape of
multipoles (phases) are uncorrelated.

Using Equation (\ref{eq_sfull}), the \cite{lambda} Maximum Likelihood
Estimate (MLE) from the WMAP ILC(3yr) map for 
the angular power spectrum  yields $S_{\rm full}^{\rm trunc, MLE} =
29431 \mu{\rm K}^4$. Compared to the value of $136670 \mu{\rm K}^4$  from the
$\Lambda$CDM best-fit to WMAP(3yr) data, 
this is not significantly unexpected, with an exclusion level of only 
$92.1\%$~C.L. 

\begin{figure}[t]
\includegraphics[height=2.7in,angle=0]{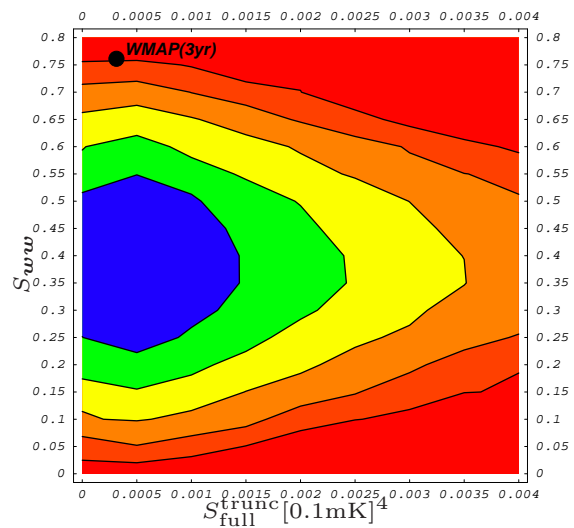}
\caption{Contour of the scatter of intrinsic alignment (\ref{eq_sww})
versus full-sky power squared (\ref{eq_sfull}). The shape can be
understood from the folding of the two respective distributions. The
total number of Monte Carlo points is $N=10^5$. The measured data point 
from WMAP 
three-year data is included. The maximum of likelihood requires 
$S_{{\boldsymbol w}{\boldsymbol w}}$ far smaller than obtained from
ILC(3yr). Consistency with the data can be excluded at $99.95\%$~C.L.
Contours correspond to lines of $1/2^n$ times the maximal
likelihood, with $n= 1,\dots, 5$.}
\label{contur1}
\end{figure}

Now we want to check for correlations between the all-sky multipole
power and the multipole alignment. As for reasons explained in the next section
we prefer the $S_{{\boldsymbol w}{\boldsymbol w}}$ statistic to
$S_{{\boldsymbol n}{\boldsymbol n}}$  in the following correlation
analysis. In Figure \ref{contur1} the scatter plot of $S_{{\boldsymbol
w}{\boldsymbol w}}$ against $S_{\rm full}^{\rm trunc}$ is shown. The form of the
contour can be understood as just the folding of the $\chi^2$-like form
of the distribution for $S_{\rm full}^{\rm trunc}$ with the gaussian-like form of
the $S_{{\boldsymbol w}{\boldsymbol w}}$ distribution. At first glance
we see from Figure \ref{contur1} that the MLE from WMAP(3yr) $S_{\rm
full}^{\rm trunc, MLE}=29431\mu$K$^4$ requires the alignment statistic to be of
middle values (around $0.4$), which is inconsistent with the
respective measured anomalous value from ILC(3yr). Moreover the lack
of any linear behaviour in the contour suggests that there is no
correlation between the two statistics.

\begin{table}[t]
\label{tab1}
\begin{ruledtabular}
\begin{tabular}{ccc}
sample size $N$ & joint $p$ & error $\Delta$ \\
\hline
100000 & $0.048 \%$ & $0.008 \%$ \\
\hspace{0.3mm} 100000\footnote{With respect to WMAP(1yr) ILC data} &
$0.001 \%$ & $0.002 \%$ \\ 
10000 & $0.02 \%$ & $0.02 \%$\\
1000 & $0 \%$ & $0.02 \%$ \\
\end{tabular}
\end{ruledtabular}
\caption{Joint likelihoods (\ref{eq_fac}) for $S_{\rm
full}^{\rm trunc}$ and $S_{{\boldsymbol w}{\boldsymbol w}}$ being in 
accordance with data simultaneously. The experimental values
refer to WMAP's ILC(3yr) map \cite{lambda} except for the second
row. The error $\Delta$ of the factorisation in equation (\ref{eq_fac}) is the
difference between left hand side and right hand side in that equation.}  
\end{table}

Given that no correlation is present between
$S_{{\boldsymbol w}{\boldsymbol w}}$ and $S_{\rm full}^{\rm trunc}$,
we would expect that the \textit{joint probability} that 
both power and alignment are in accordance with data factorizes
according to:
\begin{eqnarray}
 p\left(S_{\rm full}^{\rm trunc} \le {\rm data}\,\wedge\,S_{{\boldsymbol
w}{\boldsymbol w}} \ge {\rm data} \right) = \hspace{1in}  \nonumber \\
p_1\left(S_{\rm full}^{\rm trunc} \le {\rm data}\right) \; p_2\left(S_{{\boldsymbol
w}{\boldsymbol w}}\ge{\rm data}\right) \, .
\label{eq_fac}
\end{eqnarray}

But in reality we can only access finite statistical samples of these
quantities and the factorisation will not be exact. However we will
check the validity of (\ref{eq_fac}) within our statistical ensemble. 
When using the full sample with $N=10^5$ respectively we obtain a
\textit{joint likelihood} of $p\simeq0.05\%$. The error $\Delta$ of the factorisation,
which we define as the difference between the left hand
side in (\ref{eq_fac}) and the right hand side, is of the order
$\mathcal{O}(10^{-5})$, that is of the order of the Monte Carlo noise.
In order to track the evolution of the error $\Delta$ we also compute
the joint likelihood (\ref{eq_fac}) for smaller subsamples; see Table
\ref{tab1}. Reducing $N$ to
$N=10^4$ we obtain an even smaller joint likelihood of $p=0.02\%$ but
with an error that is of the same magnitude. With $N=10^3$ we do not
have a single hit for the joint Monte Carlos leading to $p=0\%$ with
the same error as in the $N=10^4$ case of $\Delta=0.02\%$. Note that
just one Monte Carlo hit in favor of the joint case would raise the
error here to $\Delta=0.08\%$. In the end, the convergence of the
joint likelihood appears to be very slow with respect to the sample
size $N$. 

Furthermore we are interested in the stability of the results for
$\Delta$ with respect 
to changes in the measured data. For this we choose the WMAP(1yr) values:
$S_{\rm full}^{{\rm trunc, pseudo\text{-}}C_\ell}=10154\mu$K$^4$ and
$S_{{\boldsymbol w}{\boldsymbol w}}^{\rm ILC(1yr)} = 0.7731$. We use a
sample of the full size $N=10^5$ and obtain a joint likelihood with
respect to the one-year data of $p=0.001\%$ with an error
$\Delta=0.002\%$. That is, with respect to one-year data both the
joint likelihood and its error are of the order of the Monte Carlo
noise. From the WMAP(1yr) data alone we could exclude the
joint case (\ref{eq_fac}) rather conservatively at $99.99\%$ C.L. 
This appears to be a stronger exclusion than the one from three-year
data. But we do not bother much about the difference because of the
different estimators that have been used by the WMAP team for the
angular power spectrum (pseudo-$C_\ell$ vs.~MLE) \cite{lambda}. 

We quote the most conservative result, namely the full sample joint
likelihood case for $S_{{\boldsymbol w}{\boldsymbol w}}$ and $S_{\rm
full}^{\rm trunc}$ with respect to the WMAP(3yr) data. Therefore we can exclude
that case at
$>99.95\%$ C.L.~with an error in the third digit after the comma lying
within the Monte Carlo error of the used sample ($N=10^5$). 

\begin{figure*}[t!]
\includegraphics[height=3.4in,angle=270]{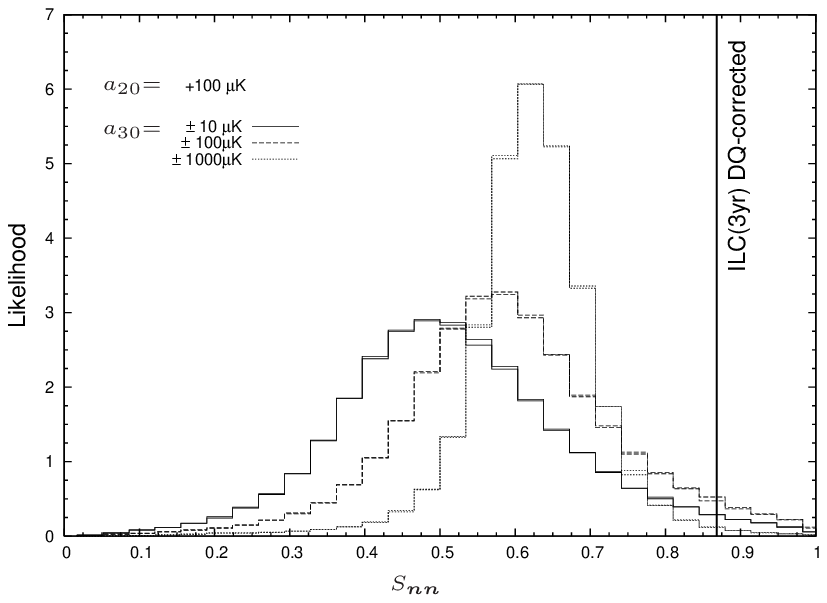}
\hspace*{8pt}
\includegraphics[height=3.4in,angle=270]{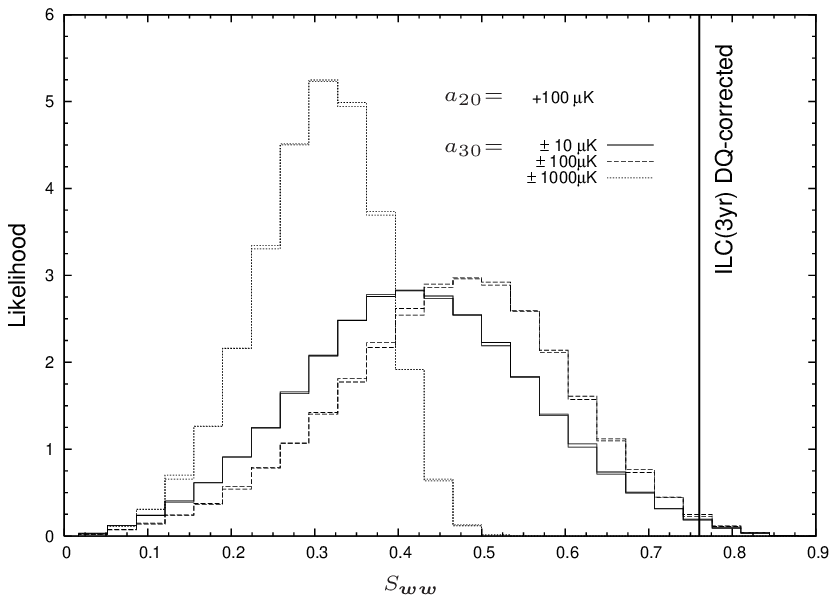}
\caption{ 
The sign of additional axial contributions $a_{\ell 0}$ has no
physical effect on the statistics $S_{{\boldsymbol n}{\boldsymbol n}}$
and $S_{{\boldsymbol w}{\boldsymbol w}}$ . For the quadrupole this
follows from the symmetry of the Legendre Polynomial $P_2$ [see
Equation (\ref{eq_alm})]. The quadrupole contribution is kept fixed at
$a_{20}=100\mu$K while the axial contribution to the octopole is
varied both in magnitude and in sign. Respective pairs of $\pm a_{3
0}$ histograms lie virtually on each other and their statistics are
thus indistinguishable. The reference histograms following from the
axially unmodified $\Lambda$CDM model (bold histograms in Figure
\ref{fig_Snn_Sww}) lie nearly on top of the displayed $a_{2
0}=100\mu$K and $a_{3 0}=\pm10\mu$K cases, and are thus not shown.}
\label{fig_asym}
\end{figure*}

Finally we analyze the correlation of the all-sky power statistic
$S_{\rm full}^{\rm trunc}$ and the intrinsic multipole alignment $S_{{\boldsymbol
w}{\boldsymbol w}}$ by quantitative means: 
 
It is well known from statistics, that when checking a finite
two-dimensional sample for correlations, the \textit{empiric
covariance} 
\begin{eqnarray}
 {\rm cov}[\,S_{\rm full}^{\rm trunc}\, ,\,S_{{\boldsymbol w}{\boldsymbol w}}\,] \equiv
\hspace{1.8in} \nonumber \\
\frac{1}{N-1} \sum_{i=1}^N \left( S_{\rm full,\,i}^{\rm trunc} - \bar{S}_{\rm
full}^{\rm trunc}  \right)\left( S_{{\boldsymbol w}{\boldsymbol w},\,i} -
\bar{S}_{{\boldsymbol w}{\boldsymbol w}} \right) 
\end{eqnarray} 
is a crucial quantity. The bar stands for the mean of a
variable. As the covariance is a scale dependent measure,
i.e.~depending on the magnitudes of the sample values $S_{{\boldsymbol
w}{\boldsymbol w},\,i}$ and $S_{{\boldsymbol w}{\boldsymbol w},\,i}$,
the dimensionless \textit{Bravais-Pearson} coefficient or
\textit{empirical correlation coefficient} is the better expression to
use:
\begin{eqnarray}
 \rho_{S_{\rm full}^{\rm trunc}\, ,\,S_{{\boldsymbol w}{\boldsymbol w}}} \equiv
\frac{\rm cov[\,S_{\rm full}^{\rm trunc}\, ,\,S_{{\boldsymbol w}{\boldsymbol
w}}\,]}{\sqrt{{\rm cov}[\,S_{\rm full}^{\rm trunc}\, ,\,S_{\rm
full}^{\rm trunc}\,] \, {\rm
cov}[\,S_{{\boldsymbol w}{\boldsymbol w}}\, ,\,S_{{\boldsymbol
w}{\boldsymbol w}}}\,]} \; . \nonumber \\
\end{eqnarray}
Finally, employing the WMAP(3yr) data we obtain an empirical correlation
coefficient of $\rho_{S_{\rm full}^{\rm trunc} , S_{{\boldsymbol w}{\boldsymbol
w}}}=-0.0027$ with respect to the full sample $N=10^5$, which indeed
indicates only marginal correlation.

\section{Inclusion of an Axis}
\label{sec4}

Now we ask what happens when introducing axial contributions on top
of a statistically isotropic and gaussian microwave sky. 
The presence of a preferred direction with axisymmetry in the CMB will
exclusively excite the zonal modes 
in case the axis is collinear to the ${\boldsymbol z}$-axis. Here we
do not bother about external directions since the internal alignments
are independent of these. Therefore such an axis will manifest
itself through additional contributions $a_{\ell 0}$. We are
considering the quadrupole and the octopole and the question arises,
in how far the sign of the axial contributions $\pm a_{\ell 0}$ plays
a role. The coefficients $a_{\ell m}$ can be reconstructed from
\begin{equation}
 a_{\ell m} = \int \frac{\Delta T}{T}(\theta, \varphi)\; Y_{\ell m}^*\;
{\rm d}\Omega \; .
\label{eq_alm}
\end{equation}
Obviously, within the quadrupole the sign of $\pm a_{2 0}$ is irrelevant
because of the symmetry of the Legendre Polynomial $P_2$ with
respect to $\theta=90^\circ$. The Legendre Polynomial $P_3$ however is
antisymmetric with respect to $\theta=90^\circ$. Therefore the
relevance of the sign of the octopole contributions $a_{3 0}$ has to
be clarified. Consequently we have chosen a fixed value for the axial
quadrupole contribution $a_{2 0}$ and have then varied the according
octopole contribution in sign and in magnitude. The results
are displayed in Figure \ref{fig_asym}.~Apparently the
$S_{{\boldsymbol n}{\boldsymbol n}}$ and $S_{{\boldsymbol
w}{\boldsymbol w}}$ statistics that are important here, do not
distinguish between the sign of the applied axial effect. Therefore we
need not to bother about the signs of the $a_{\ell 0}$ and let them
henceforth be positive.

In Figure \ref{fig_Snn_Sww} the evolution of the $S_{{\boldsymbol w}{\boldsymbol
w}}$ and $S_{{\boldsymbol n}{\boldsymbol n}}$ statistics with respect
to increasing axial contributions is displayed in terms of likelihood histograms:

Let us first look at the evolution of
the $S_{{\boldsymbol n}{\boldsymbol n}}$ statistic. This expression
measures the average $|\cos|$ of the angles between the quadrupole oriented
area and the octopole areas. The pure Monte Carlo peaks at $0.5$
reflecting the fact that the average distance of four isotropically
distributed vectors on 
a half--sphere from each other is $60^\circ$ in the case of
statistical isotropy. It is a half-sphere because the signs of the
multipole vectors are arbitrary and so we choose them all to point to
the northern hemisphere. When increasing the contribution of the axial
effect the multipoles become increasingly zonal and arrive at being 
purely zonal in a good approximation at values of $a_{\ell
0}=1000\mu$K. On the level of the multipole vectors this means that
their cross products all move to the equatorial plane (see
Figure \ref{fig_mollw}). That is the
reason why the histogram in Figure \ref{fig_Snn_Sww} (left) moves to the right
when we increase the axial effect, because now
isotropy is broken from the half-sphere to the half-circle making the
$S_{{\boldsymbol n}{\boldsymbol n}}$ histogram peak sharper at higher values. 
The measured value from the ILC(3yr) map of $S_{{\boldsymbol
n}{\boldsymbol n}}^{\rm ILC(3yr)}=0.868$ is anomalous at $98.3\%$
C.L.~with respect to the pure Monte Carlo (bold histogram in
Figure \ref{fig_Snn_Sww}) which stands for the statistically isotropic
and gaussian model. By adding axial contribution the maximal
improvement is reached at $a_{\ell 0}=100\mu$K where the
ILC(3yr) becomes unexpected at $96.7\%$ C.L. Further enhancement of the
axial effect makes the $S_{{\boldsymbol n}{\boldsymbol n}}$ statistic
more and more narrow around an expectation value $<0.7$. This makes it impossible to
remove the anomaly in the $S_{{\boldsymbol n}{\boldsymbol n}}$
cross-alignment with respect to the ILC(3yr) experimental value
only by increasing the axial contribution to high enough values.

\begin{figure}
\includegraphics[height=2.7in,angle=0]{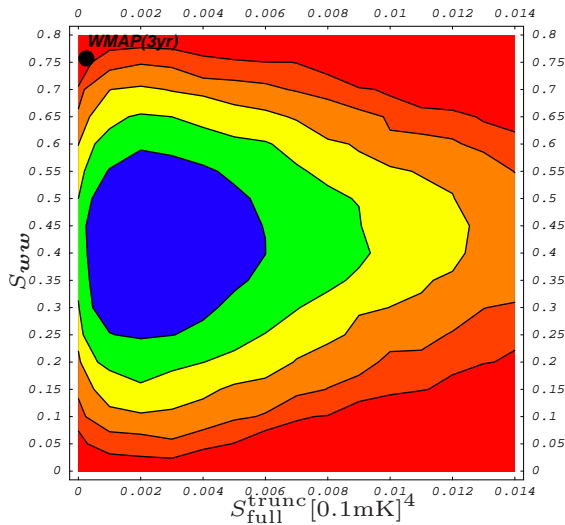}
\caption{ 
Scatter contour of pairs of $S_{{\boldsymbol w}{\boldsymbol
w}}$ and $S_{\rm full}^{\rm trunc}$ after an axial modification of $a_{\ell
0}=70\mu$K has been applied; this is the contribution involving
maximal improvement in $S_{{\boldsymbol w}{\boldsymbol w}}$ (see
Figure \ref{fig_Snn_Sww}). The total number of Monte Carlo pairs is $N=10^5$.
Note that the horizontal axis now runs from zero to $1.4\times 10^{-6}{\rm
mK}^4$, whereas in Figure \ref{contur1} the maximal displayed value is
$4\times 10^{-7}{\rm mK}^4$.
The inclusion of a preferred axis leaves all-sky
multipole power and intrinsic alignment totally uncorrelated and
inconsistent with the WMAP(3yr) data. Contour lines are defined as in
Figure~\ref{contur1}.} 
\label{contur2}
\end{figure}

On the other hand the $S_{{\boldsymbol w}{\boldsymbol w}}$ statistic
additionally measures the modulus of the sin of the angles between 
the multipole vectors themselves. As can be seen from Figure~\ref{fig_mollw}
multipole vectors are all moving toward the north pole
clustering more and more as the axial contribution is enhanced.
The $S_{{\boldsymbol w}{\boldsymbol w}}$ statistic measures 
the average of the modulus of the products of the sin of angles
between quadrupole vectors, octopole vectors and the cos of the angle
between the area vectors. Therefore on top of the information already
contained in $S_{{\boldsymbol n}{\boldsymbol n}}$ the $S_{{\boldsymbol
w}{\boldsymbol w}}$ statistic is able to go to zero for highest zonal
contamination as the closeness of the multipole vectors in that case
dampens the product of sines and cosines quadratically to arbitrary small
values. Thus we find that $S_{{\boldsymbol w}{\boldsymbol w}}$ is
the more convenient statistic for further analyses, as it
does contain more information than the $S_{{\boldsymbol n}{\boldsymbol
n}}$ statistic and additionally shows a simple and clear asymptotic
behavior. In the case of this statistic the anomaly is significant at 
$99.5\%$C.L.~with respect to $S_{{\boldsymbol w}{\boldsymbol w}}^{\rm 
ILC(3yr)}=0.7604$. Similarly to before the maximal improvement is
reached with an axial contribution of $a_{\ell 0}=70\mu$K, which
degrades the anomaly in $S_{{\boldsymbol w}{\boldsymbol w}}$ to
$99.2\%$C.L.  

Now we return to the correlation analysis of the alignment with the
pure multipole power $C_\ell$.
When introducing an axial effect, say $a_{\ell 0}=100\mu$K, we
improve the fit to the $S_{{\boldsymbol w}{\boldsymbol w}}$ statistic, but
interestingly the multipole power anomaly becomes much more 
pronounced. This behaviour is expected \cite{rse1,rse2} for the
$C_\ell$-distribution (being a modified $\chi^2$-distribution) when
the axial contribution is enhanced, but it is unexpected that
exactly the same happens for a multipole power distribution `that
knows of the intrinsic alignment of quadrupole and octopole'. This
indicates that there is no correlation at all between multipole power
and the phase alignment even when they are tuned to each other.

\begin{figure}
\includegraphics[height=2.7in,angle=0]{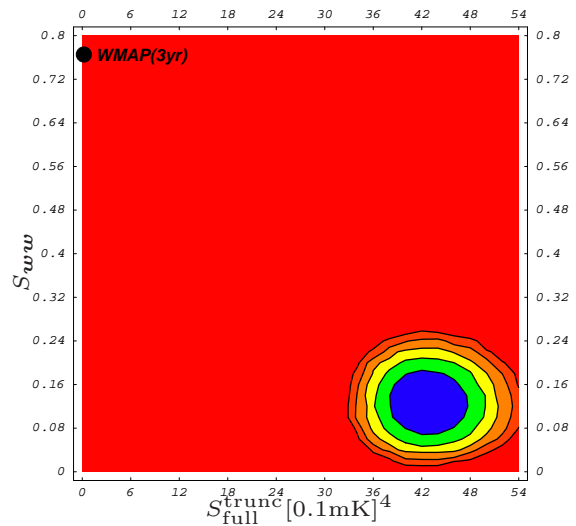}
\caption{Contour plot of the scatter of pairs $(S_{{\boldsymbol
w}{\boldsymbol w}},S_{\rm full}^{\rm trunc})$ after a strong axial contingent of 
$a_{\ell 0}=1000\mu$K is induced to the multipole vectors (see also 
Figures \ref{fig_mollw},\ref{fig_Snn_Sww}). The total number of Monte
Carlo pairs 
is $N=10^5$. The all-sky power statistic reacts heavily as the scale
on the $S_{\rm full}^{\rm trunc}$-axis is shifted by four orders of magnitude
with respect to the case of $a_{\ell 0}=70\mu$K
(Figure \ref{contur2}). The likelihood maximum departs very
articulately from the WMAP(3yr) data. The contour lines are defined
like in Figure~\ref{contur1}.} 
\label{contur3}
\end{figure}

Proceeding with the analysis of correlations between alignment and the
full-sky power statistic, again we try to provoke correlation with
the help of axial symmetry in the CMB. In fact we apply an axial
effect of the ideal magnitude ($a_{\ell 0}=70\mu$K) in order achieve
larger values in $S_{{\boldsymbol w}{\boldsymbol w}}$. The negative
result is shown in Figure~\ref{contur2}: 
As $S_{\rm full}^{\rm trunc}$ is a linear
combination of squared $C_\ell$ distributions it is a sharply peaked
$\chi^2$-like distribution being very sensitive to axial
contributions. Therefore the contour in Figure~\ref{contur3} is fairly
shifted to the right (to higher values in $S_{\rm full}^{\rm trunc}$) and
broadened with respect to the axially unmodified case, obviating any
correlation with the intrinsic alignment. The shape of the overall
contour is  roughly left invariant by the scale shift in $S_{\rm
full}^{\rm trunc}$.   

Figure \ref{contur3} illustrates the pure zonal case. Here a whole
$a_{\ell 0}=1000\mu$K has been induced into the multipole vectors. 
Again, due to the sensitivity of $S_{\rm full}^{\rm trunc}$ to axial
contamination this pushes the allowed region in the scatter plot
to very high values in full-sky power squared, degenerating the
contour to a `small' area far away from the measured three-year WMAP
values. No change in correlation is observable. 

Obviously, no coupling of the multipole power statistic
and the intrinsic alignment can be driven in favor of the anomalous
experimental CMB data by an additional axisymmetric effect on top
of the primordial fluctuations.

\section{Conclusions}
\label{sec5}

We have shown that a literal interpretation of the `Axis of Evil' as an 
axisymmetric effect is highly incompatible with the observed microwave 
sky at the largest angular scales. The
formalism of multipole vectors was used to separate directional
information from the absolute power of multipoles on
the CMB sky. Considered were two choices of statistic, measuring
the intrinsic cross-alignment between the quadrupole and octopole: the
$S_{{\boldsymbol n}{\boldsymbol n}}$ and the $S_{{\boldsymbol
w}{\boldsymbol w}}$ statistic. We confirm that the $S_{{\boldsymbol
w}{\boldsymbol w}}$ statistic contains more information on the
multipoles and that it has more discriminative power as an axial effect is
included. The presence of an axial symmetry
in the CMB would excite zonal modes which are, in the
frame of the axis, additional $a_{\ell
0}$ contributions in the language of the harmonic decomposition.
Both statistics
($S_{{\boldsymbol n}{\boldsymbol n}}$ and $S_{{\boldsymbol
w}{\boldsymbol w}}$) reach slightly better agreement with the measured 
values from the ILC(3yr) map at amplitudes of roughly
$a_{\ell 0}=100\mu$K. Further enhancement of the axial effect
only reduces consistency with WMAP(3yr) data.

Especially we have assayed in what way the alignment anomaly between 
quadrupole and octopole can affect the respective multipole power. We
made several tests where we identified and selected the `anomalous
$a_{\ell m}$' that are still consistent with data and checked whether
the resulting distribution from these $a_{\ell m}$ for either power or
alignment shows any change with respect to the unbiased case. For the
all-sky multipole power we make use of the statistic $S_{\rm
full}^{\rm trunc}$.
We demonstrated that the correlation between $S_{\rm full}^{\rm trunc}$
and intrinsic alignment is only marginal (correlation coefficient of
$-0.0027$). Thus a factorisation of the probability for the joint case
into a product of the respective probabilities is allowed.

We argued that the combined case
of the measured all-sky power and the quadrupole-octopole alignment
is anomalous  at $>99.95\%$ C.L. with respect to the WMAP three-year data. 
The correlation picture leaves no space for an axisymmetric effect in
the large-angle CMB. 

These findings complement our previous studies \cite{rse1} of the 
interplay of an axisymmetric effect and the extrinsic CMB anomalies 
(correlation with the motion and orientation of the Solar system 
\cite{schwarz04}). In that work it was shown that an axisymmetric 
effect might help to explain a Solar system alignement. Finally, 
this study rules out that possibility.

But there is a loophole. Here and in \cite{rse1} 
we only considered \textit{additive}
modifications of the $a_{\ell m}$. Still, a preferred axis could also
induce \textit{multiplicative} modifications in all $a_{\ell m}$
\cite{huhu06}. This could avoid the problem of additional multipole
power. However, multiplicative effects could only be achieved  
by non-linear physics, like systematics of the measurement or the 
map making process.

A modelling that would be able to consistently remove
both the power and the intrinsic alignment problem for low-$\ell$
must mobilize a more complex pattern of modifications than the one
induced by an axisymmetric effect. As already indicated by e.g.~the odd
extrinsic alignment with the ecliptic
\cite{schwarz04,copi05,copi06,rse1,rse2} the CMB anomalies do rather
require a special plane than a preferred axis. The so called `Axis of
Evil' appears as just the normal vector of that plane, but no
axial symmetry is present within that plane. 

\begin{acknowledgments}
\noindent We thank Dragan Huterer for pointing out the importance of
looking at the cross-correlations and Glenn Starkman
for important discussions and suggestions regarding the presentation, 
as well as Amir Hajian and David Mota for useful comments.
We want to thank the referee for useful comments improving
the paper and Bastian Weinhorst for advice with \texttt{MATHEMATICA}.
We acknowledge the use of the Legacy Archive
for Microwave Background Data Analysis (LAMBDA) provided by the NASA
Office of Space Science. The work of AR is supported by the DFG under
grant GRK 881.
\end{acknowledgments}

\end{document}